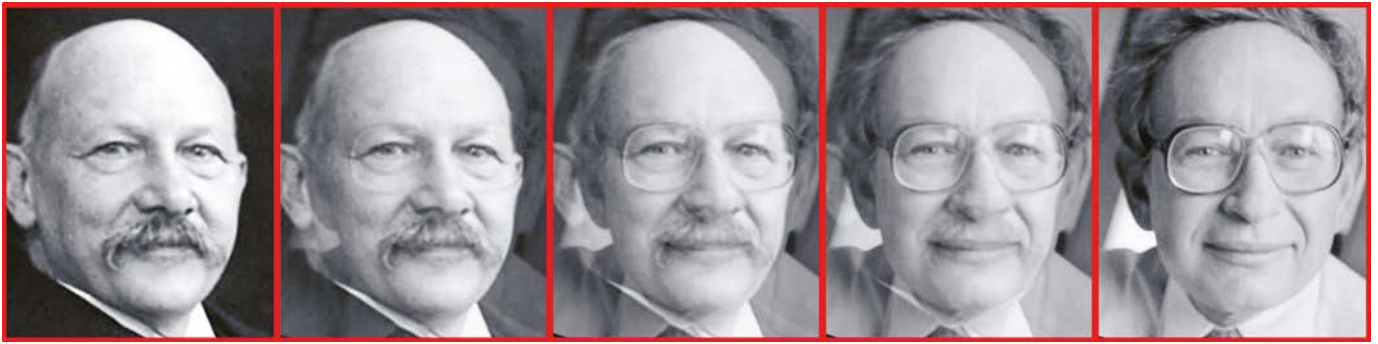

# THE MANY FACES (PHASES) OF STRONG CORRELATIONS

■ Silke Paschen[1] and Qimiao Si[2] – DOI: https://doi.org/10.1051/epn/2021407
■ [1] Institute of Solid State Physics, TU Wien, Austria
■ [2] Department of Physics and Astronomy, Rice Center for Quantum Materials, Rice University, USA

There has been considerable recent progress in discovering and understanding quantum phases and fluctuations produced by strong correlations. Heavy fermion systems are an ideal platform for systematic studies because low and competing energy scales make them highly tunable. As such the phases (faces) of strong correlations transform continuously into one another.

▲ Morphing images of the faces of Nobel laureates Heike Kamerlingh Onnes (1913) and Philip Warren Anderson (1977).

Mobile electrons in simple metals can be reasonably well described in a free electron picture – the Sommerfeld model of a Fermi gas. At the forefront of current research, however, are materials in which the electron-electron Coulomb interaction $U$ is sizable and can not be ignored. Such "strongly correlated" materials present a wealth of phenomena that are intriguing for fundamental studies and promising for electronic applications.

### Strongly correlated electrons

In theoretical descriptions, $U$ is compared to an energy scale that relates to the kinetic energy of the electrons, typically the hopping energy $t$ or the bandwidth $W$. When $U/W$ is small, the low-energy physics of the many-electron system is described in terms of noninteracting conduction electrons, with the correlations serving as a perturbation. In other words, the conduction electrons are the only building blocks of the low-energy physics. When $U/W$ reaches or exceeds order unity, the correlations produce new degrees of freedom – such as spins – that serve as the effective building blocks for the low-energy physics.

This applies to strongly correlated materials as exemplified by the high-temperature copper- and iron-based superconductors and heavy fermion metals. A key characteristic of these systems is that the correlations produce a plethora of quantum phases. Viewed in terms of the effective building blocks, the electron correlations are more explicitly manifested as competing interactions that promote different kinds of ground states. For example, in the 4$f$-electron-based heavy fermion metals, the $f$ electrons act as localized moments of effective spin $S = 1/2$. The spins interact with a separate band of $spd$ electrons through an antiferromagnetic Kondo interaction. They also couple to each other by an RKKY spin-exchange interaction, produced by the spin polarization of the conduction electrons, that is typically antiferromagnetic as well. The competing tendencies of the two interactions, to promote inter-species Kondo singlets and inter-spin singlets, respectively, lie at the heart of the heavy fermion physics.

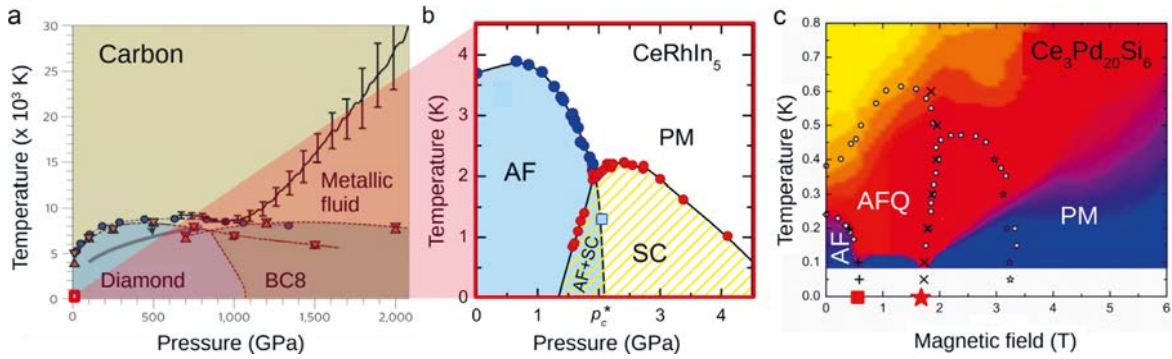

**FIG. 1:** Temperature-tuning parameter phase diagrams of a non-interating material (a, carbon under ultra-high pressure, BC8 is an expected metastable phase [1]) and two strongly correlated materials (b, CeRhIn$_5$ [2] and c, Ce$_3$Pd$_{20}$Si$_6$ [3]). The tunability is vastly enhanced in strongly correlated materials as seen from a comparison of the scales of panels a and b. Red colour in c represents linear-in-temperature resistivity.

### Quantum criticality

In many of these materials non-thermal control parameters such as pressure or magnetic field can tip the balance between the different interactions. As a consequence, a given phase can be suppressed and another one stabilized as function of a control parameter (Fig. 1). When this process occurs continuously down to the absolute zero in temperature, the ensuing zero-temperature *quantum* phase transition is a quantum critical point (QCP). It separates distinct phases, for instance an antiferromagnet (AF) from a paramagnet (PM). The QCP anchors quantum critical fluctuations which can strongly affect the finite temperature properties of the material. A well-known example are electrical resistivities with unusual temperature dependences, most notably the linear-in-temperature "strange metal" behaviour. This is seen to emerge from two magnetic-field tuned QCPs in the heavy fermion compound Ce$_3$Pd$_{20}$Si$_6$, at the border of phases with AF and antiferroquadrupolar (AFQ) order, respectively (Fig. 1c). Another intriguing effect is that quantum critical fluctuations can stabilize new phases. For instance, in the heavy fermion compound CeRhIn$_5$, a phase of unconventional superconductivity (SC) is stabilized as an AF phase is continuously suppressed to zero by pressure (Fig. 1b).

An overall understanding of heavy fermion quantum criticality comes in the form of a global phase diagram (Fig. 2). The Kondo interaction $J_K$ promotes quantum fluctuations. The ensuing Kondo effect leads to a ground state in which the local moments form a spin singlet with the conduction electrons. It converts the local moments into electronic excitations, the composite heavy fermions (Box). The latter hybridize with the conduction electrons, and the Fermi surface expands and is called large (subscript L in Fig. 2). At smaller $J_K$, the local moments establish antiferromagnetic order from the dominating RKKY interaction. They form spin singlets among themselves, which destabilizes the inter-species Kondo singlets. With this Kondo destruction [3], the Fermi surface is formed by the conduction electrons alone and is called small (subscript S in Fig. 2). The RKKY interaction also promotes quantum fluctuations, which can be enhanced by geometrical frustration $G$. The global phase diagram captures the effect of both types of quantum fluctuations. Importantly, it goes beyond the Landau framework where quantum critical fluctuations derive uniquely from the suppression of the order parameter as the broken-symmetry phase (AF in our case) gives way to a paramagnetic phase (P). Here, in addition to the Landau order parameter, the Kondo effect and its destruction characterize the quantum phases and fluctuations.

Experimentally, the change of the Fermi surface across such a Kondo destruction QCP was evidenced by Hall effect measurements [3]. Associated with this transition is the strange metal linear-in-temperature electrical resistivity referred to above. In addition, a recent optical conductivity investigation in the THz range – appropriate to probe the low energy scales of heavy fermion systems – revealed quantum critical charge fluctuation. The result indicates that fermionic degrees of freedom, in addition to the bosonic order parameter fluctuations, govern the quantum criticality [4]. This is exactly the behaviour expected in a Kondo destruction QCP. Similar phenomena have also been observed in other classes of strongly correlated electron systems, most notably the high-temperature cuprate superconductors [5,6]. The common features raise the exciting possibility that high-temperature superconductivity is an emergent phase stabilized by the type of quantum criticality that goes beyond the order-parameter-fluctuation paradigm.

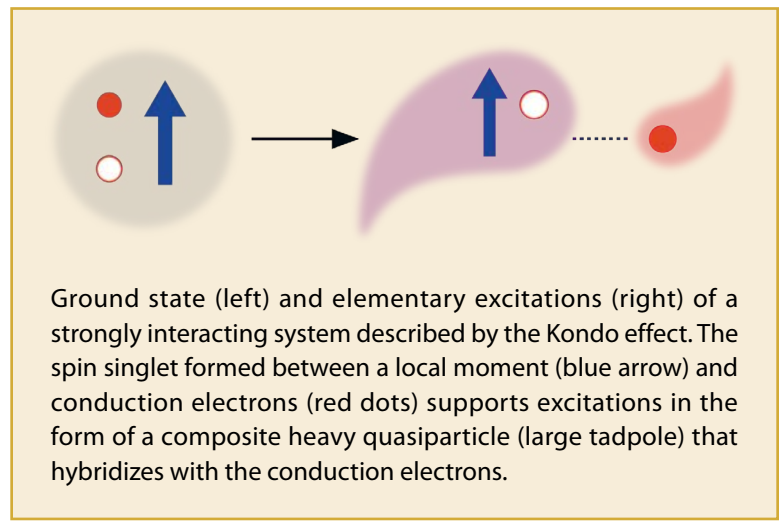

Ground state (left) and elementary excitations (right) of a strongly interacting system described by the Kondo effect. The spin singlet formed between a local moment (blue arrow) and conduction electrons (red dots) supports excitations in the form of a composite heavy quasiparticle (large tadpole) that hybridizes with the conduction electrons.

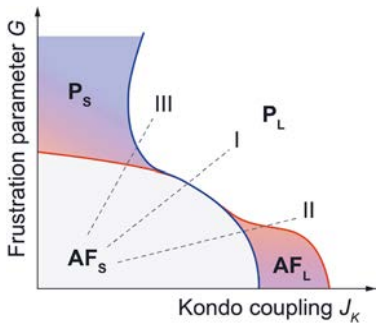

◀ **FIG. 2:** Theoretical phase diagram of AF Kondo lattice models [3]. The different quantum phases are described in the text. The paths I, II and III represent different tuning trajectories that realize distinct sequences of quantum phase transitions and the associated QCPs. This global phase diagram provides a framework to guide the exploration of heavy fermion compounds with different spatial dimensionality or lattice geometry, which may realize the different regimes of the parameter space.

### Correlation-driven electronic topology

Another area of recent developments concerns the combination of strong electron correlations and nontrivial electronic topology. Topological states are of interest because they are imprinted in a material for symmetry reasons. As such they are expected to be robust against disorder. This makes topological materials candidates for new electronic devices. Topological phases are inherently not characterized by a (Landau) order parameter. To delineate their boundaries and devise global phase diagrams as described above for (topologically trivial) strongly correlated electron systems requires to produce correlated topological states, identify their characteristics, and tune them by external parameters.

An interesting recent discovery is that strong correlations can lead to giant topological responses. A new state of matter, dubbed Weyl-Kondo semimetal, was shown to exhibit thermodynamic and electrical transport signatures that overshoot expectations for their non-interacting counterparts by orders of magnitude [3] (Fig. 3). In addition, topological features in strongly correlated systems are much more readily controllable by external parameters than in the non-interacting case. These two features will help to identify and tune topological phases and thus map out their global phase diagrams. This in turn will allow to reveal the underlying principles. Whether quantum critical fluctuations can also promote emergent topological phases – in analogy with unconventional superconductivity for topologically trivial systems – is an exciting question for future studies. The observation of quantum criticality, with characteristics of Kondo destruction, in a candidate Weyl-Kondo semimetal certainly nourishes this hope [7].

### Outlook

We have only been able to give a few examples of the considerable recent progress in the field of strongly correlated quantum materials. Other directions that are actively being explored include the investigation of phases governed by building blocks beyond simple spins and conduction electrons, enhanced frustration or hybrid interactions. In addition to the multitude of strongly correlated materials classes, artificial structures such as twisted bilayer systems are now available and hold promise for studies of correlation physics via moiré potentials. For further reading we refer the reader to [3]. Finally, new experimental techniques, for instance measurements at ultralow temperatures [9] or the growth of quantum critical heavy fermion compounds by molecular beam epitaxy [4], open entirely new possibilities. One of the grand challenges to attack is to exploit the rich physics of correlated quantum materials for quantum devices. With their strongly amplified responses, the prospect is certainly high. ∎

### About the Authors

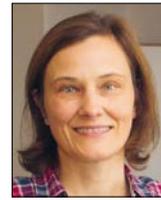

**Silke (Bühler-)Paschen** is an experimental condensed matter physicist and professor at TU Wien. She is APS fellow and recipient of an ERC Advanced Grant. She and her team study strongly correlated quantum materials with a wide arsenal of techniques – from the synthesis of bulk and thin-film single crystals to advanced measurements into the microkelvin regime.

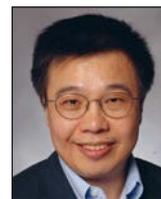

**Qimiao Si** is a theoretical condensed matter physicist and professor at Rice University. He is a fellow of APS and British IOP and recipient of a Humboldt Research Award. His research tackles theoretical models with the aim of uncovering and advancing organizing principles that may be universal across strongly correlated electron systems.

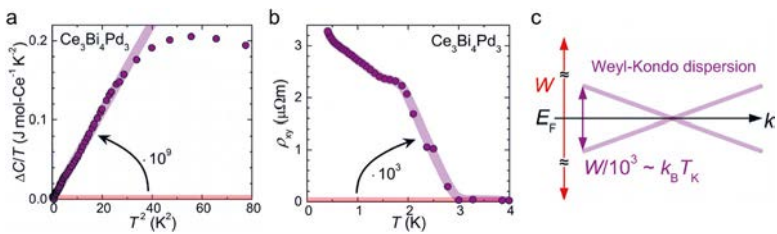

▲ **FIG. 3:** Giant topological responses of the Weyl-Kondo semimetal $Ce_3Bi_4Pd_3$ [3,8]. The giant slope of the electronic specific heat coefficient $\Delta C/T$ as function of $T^2$ (a) derives from linearly dispersing electronic bands, the width of which is strongly renormalized by the Kondo effect (characterized by the Kondo temperature $T_K$) with respect to the width of the underlying non-interacting band $W$ (c). The giant spontaneous Hall resistivity $\rho_{xy}$ (b) is attributed to Berry curvature divergences at the Weyl nodes which, in a Weyl-Kondo semimetal, are situated in close vicinity to the Fermi level. Electronic transport is therefore very strongly affected by the associated fictitious magnetic monopoles in momentum space.